\documentclass[twocolumn,showpacs,preprintnumbers,nofootinbib,prd,superscriptaddress,groupedaddress,10pt]{revtex4-1}

\expandafter\let\csname equation*\endcsname\relax
\expandafter\let\csname endequation*\endcsname\relax

\usepackage{graphicx,amssymb,amsmath,amsthm,amsfonts,epsfig}
\usepackage[colorlinks=true]{hyperref}
\usepackage[usenames]{color}
\usepackage{epstopdf}

\usepackage{bm}
\usepackage{dcolumn}
\usepackage[utf8]{inputenc}
\usepackage{latexsym}
\usepackage{rotating}
\usepackage{xcolor}
\usepackage{longtable}
\usepackage{enumerate}
\usepackage{mathtools}
\usepackage{url}
\newcommand{\ben}{\begin{enumerate}}
\newcommand{\een}{\end{enumerate}}

\def\be{\begin{equation}}
\def\ee{\end{equation}}
\def\bea{\begin{eqnarray}}
\def\eea{\end{eqnarray}}
\newcommand{\beq}{\begin{eqnarray}}
\newcommand{\eeq}{\end{eqnarray}} 
\newcommand{\ba}{\begin{align}}
\newcommand{\ea}{\end{align}}

\def\M{\mathcal{M}}

\def\ba{\bar{a}}

\newcommand{\tn}{\textnormal}

\makeatletter
\renewcommand\footnoterule{%
  \kern-3\p@
  \hrule\@width2.5cm
  \kern2.6\p@}
\makeatother

\usepackage{cleveref}
\crefname{subsection}{subsection}{subsections}
\crefname{section}{\S}{\S\S}
\Crefname{section}{\S}{\S\S}

\def\GB{\mathrm{GB}}
\def\ppE{\mathrm{ppE}}

\def\EdGB{\mathrm{EdGB}}
\def\dCS{\mathrm{dCS}}

\usepackage{mathrsfs}

\newcommand{\brackets}[1]{\left( #1 \right)}
\newcommand{\sqbrackets}[1]{\left[ #1 \right]}
\newcommand{\curbrackets}[1]{\left\{ #1 \right\}}

\begin{document}

\title{Bounding alternative theories of gravity with multiband GW observations}
\author{
Giuseppe Gnocchi$^{1}$,
Andrea Maselli$^{1}$,
Tiziano Abdelsalhin$^{1}$,
Nicola Giacobbo$^{2,3,4}$,
Michela Mapelli$^{2,3,4,5}$
}
\address{
$^{1}$Dipartimento di Fisica, Sapienza Università di Roma \& Sezione INFN Roma1, P.A. Moro 5, 00185, Roma, Italy\\
$^{2}$Dipartimento di Fisica e Astronomia ``G. Galilei'', Università di Padova, Vicolo dell'Osservatorio 3, I--35122, Italy\\
$^{3}$INAF-Osservatorio Astronomico di Padova, Vicolo dell'Osservatorio 5, I--35122 Padova, Italy\\
$^{4}$INFN-Padova, Via Marzolo 8, I--35131 Padova, Italy\\
$^{5}$Institut f\"{u}r Astro- und Teilchenphysik, Universit\"{a}t Innsbruck, Technikerstrasse 25/8, 6020 Innsbruck, Austria}
\begin{abstract}
We study the constraints on alternative theories of gravity that can be determined by multiband observations of 
gravitational wave signals emitted from binary black hole coalescences. We focus on three types of General Relativity 
modifications induced by a generalized Brans-Dicke theory, and two classes of quadratic gravity, Einstein-dilaton-Gauss-Bonnet 
and dynamical Chern-Simons. Considering a network of space- and ground-based detectors, supplied 
by a population of spinning binaries black holes, we show how the multiband analysis improves the existing bounds on 
the theory's parameters  by several orders of magnitude, for both pre- and post-Newtonian deviations. Our results also show 
the fundamental role played by an interferometer in the frequency range between LISA and advanced detectors, in 
constraining possible  deviations from General Relativity.
\end{abstract}
\maketitle

\section{Introduction}

Gravitational waves (GW) emitted by the coalescence of black hole (BH) binaries are among the cleanest and 
most valuable tools to investigate the features of gravity and to test the predictions of General Relativity (GR) 
in the highly-relativistic, strong-field regime \cite{Berti:2015itd,Berti:2018cxi,Sathyaprakash:2019yqt}. GW signals 
observed so far have been deeply analyzed by the LIGO/Virgo Collaboration \cite{TheLIGOScientific:2016src,TheLIGOScientific:2016pea,Abbott:2018lct} to find 
possible deviations from GR. Recently, a series of tests has been performed on the joint datasets collected by 
advanced detectors, showing no sign of inconsistency with Einstein's theory \cite{LIGOScientific:2019fpa}. 
Future binary black hole (BBH) observations are expected to lower statistical errors on the source's parameters, 
while numerical and semianalytical studies promise to reduce the systematics in the waveform's modeling 
\cite{Berti:2018cxi,Berti:2018vdi,Barack:2018yly}. Moreover, a new family of GW interferometers will complement the existing 
detectors with both ground-based and space facilities.
 
By the mid-2030s, the satellite LISA \cite{Audley:2017drz,Berti:2019xgr,Cutler:2019krq,McWilliams:2019fng} 
will be ready to start an observational campaign aimed to detect GW signals emitted between $10^{-4}$ and 
$10^{-1}$ Hz. With a sensitivity curve being designed to follow the evolution of supermassive black holes, 
either in symmetric or extreme mass ratio binaries, LISA will also provide a complementary window, in the low 
frequency band, for stellar-mass black holes which are among the primary targets of ground-based interferometers. 
The relevance of space and terrestrial joint detections has been recently investigated \cite{Sesana:2016ljz}, showing the 
improvement in the measurements of the source's parameters, which is crucial, as an example, for a precise binary 
localisation. Had LISA been operating during the first LIGO/Virgo observations, this would have allowed us to 
know with great accuracy the moment at which the signal would have entered the bandwidth on Earth, 
boosting the quality of data analysis~\cite{Sesana:2016ljz,Gerosa:2019dbe}. 

Multiband observations also represent a powerful approach to probe theoretical foundations of General Relativity 
and test gravity modifications \cite{Berti:2015itd} in different dynamical regimes. Stellar-mass BBH observed 
by both space- and ground-based interferometers evolve through a wide range of frequencies emitting GW in the millihertz LISA's 
band for years (the early inspiral phase), before chirping at high frequencies and producing a short signal (the 
late inspiral and merger phases) in the LIGO/Virgo band, around approximately $100$ Hz. 
A joint detection would help to constrain the 
source's parameters, which may be dominant either at low or at high frequencies, and therefore would be 
measured with different accuracy by space and terrestrial detectors. For example, a pre-Newtonian effect, i.e., 
a non-GR correction which modifies the waveform before the leading quadrupolar order, plays a major role in 
the early evolution of a binary system. Strong bounds on this modifications can be placed by observing double pulsars 
in the electromagnetic channel \cite{Gerard:2001fm,Kramer:2006nb,Bhat:2008ck}. For binary black holes, this effect 
would be more dominant in the low-frequency spectrum of LISA, leading to very different constraints with respect to 
a possible LIGO/Virgo detection \cite{Barausse:2016eii}.
 
Besides LISA and current advanced interferometers, the GW discovery is pushing the development of new facilities. 
KAGRA is close to completion \cite{Akutsu:2018axf,Somiya:2011np}, while huge improvements in the sensitivity will be 
given by third generation detectors, as the Einstein Telescope \cite{Punturo:2010zz} or 
the Cosmic Explorer \cite{Dwyer:2014fpa,Evans:2016mbw}. Finally, new space satellites are under active study, as the
Japanese B-DECIGO \cite{Nakamura:2016hna}, which is designed to bridge the gap between LISA and ground-based 
interferometers, and promise to observe both BH and neutron star binaries with exquisite precision\footnote{New, 
conceptually different detectors are also under investigation, as atomic GW interferometers, in the 
subhertz frequency band \cite{2018NatSR...814064C}.} \cite{Isoyama:2018rjb}.

This network of interferometers actually represents a  wideband detector able to measure with 
pinpoint accuracy the parameters of binary sources, and to test gravity throughout the full orbital evolution 
up to the merger phase. 
In this paper, we use such global web of interferometers to explore how multiband GW detections can  
constrain the fundamental parameters of three different alternative theories of gravity: a generalized model of the 
Fierz-Jordan-Brans-Dicke theory \cite{Jenkins:2013fya,Barausse:2016eii} and two types of quadratic gravity, namely 
Einstein-dilaton-Gauss-Bonnet \cite{Kanti:1995vq,Moura:2006pz} and dynamical Chern-Simons \cite{Jackiw:2003pm,Alexander:2009tp,Delsate:2014hba}. 
In such theories, the gravitational interaction is mediated by an extra scalar field coupled with terms proportional to the 
curvature. The presence of the scalar field activates new physical mechanisms, such as the emission of dipole radiation, which 
increase the overall gravitational wave flux emitted by binary systems and hence change their orbital evolution. 
These corrections modify the GW phase at different post-Newtonian (PN) orders and  then affect the signal 
in different frequency regimes. They represent therefore a test bed for the full potential of the multiband analysis.  

We model the GR deviations using the parametrized post-Einsteinian (ppE) approach \cite{Yunes:2009ke,Tahura:2018zuq}. Similar to the 
low-velocity, weak-field PN expansion of the metric and matter variables \cite{Blanchet:2013haa}, the ppE formalism maps 
model-independent deviations from GR directly into the GW signals emitted by binary sources. 
The effectiveness of this framework to detect deviations produced by alternative theories using synthetic and real 
data has been deeply explored so far 
\cite{Narikawa:2016uwr,Vallisneri:2013rc,Chatziioannou:2012rf,Sampson:2013lpa,Sampson:2013wia,Yunes:2010qb,Barausse:2016eii,Maselli:2016ekw,Yunes:2016jcc,Chamberlain:2017fjl,Zimmerman:2019wzo,Nair:2019iur} 
(we refer the reader to the review \cite{Yunes:2013dva} and reference therein for an extensive lecture on the subject). 
In particular, Yunes {\it et al.} have explored the fundamental physics implications that current and future GW detections 
may have on a large set of modified theories of gravity \cite{Yunes:2016jcc,Chamberlain:2017fjl}. Astrophysical bounds on the 
lowest order ppE coefficients have also been derived using observations of relativistic binary pulsars \cite{Yunes:2010qb}. 

In this scenario, the ppE formalism represents a precious tool to perform multiband analysis which spans the GW 
spectrum within the frequency range $[10^{-4}-10^3]$ Hz, i.e., for a global network composed of space- and ground- 
based interferometers. Joint constraints on scalar-tensor theories of gravity, with a $-1$PN correction in the GW phase, 
have been studied by Barausse \emph{et al.} assuming LISA and advanced LIGO/Virgo observations of prototype binaries 
\cite{Barausse:2016eii}. This work has shown how the constraints on the non-GR parameters can benefit from 
multiple detectors, improving the bounds from single measurements of various orders of magnitude.

In this paper, we pursue a similar path, and we extend previous studies in order to (i) broaden the analysis to 
different classes of alternative theories of gravity that introduce distinct ppE corrections in the waveform, (ii) consider 
an astrophysical population of spinning stellar-mass BH binaries, and (iii) take into account third generation interferometers 
such as the Einstein Telescope \cite{Hild:2010id,ETWhite} and a second space instrument in the hertz band given by the 
proposed Japanese B-DECIGO \cite{Isoyama:2018rjb}. We derive the distribution of statistical errors on theory's 
parameters, computing the projected bounds for different detector's configurations. We show how space and terrestrial 
interferometers, supplied by the population of BBHs, represent an incredible opportunity to test gravity 
modifications which affect GW signals on a wide range of regimes.

\section{Binary black hole population}
\label{sec:stellarmodels}

The astrophysical BBH population used for the injections is obtained by combining state-of-the-art population 
synthesis simulations with a cosmological simulation, as already described in \cite{mapelli2017,mapelli2018}. 
In particular, the population-synthesis simulations provide information on BBH mass and delay time (i.e., the 
time elapsed between the formation of the progenitor stellar binary and the time of the merger), while the 
cosmological simulation outputs provide information on the merger redshift and on the host galaxy of the BBH. 

We use the publicly available {\sc Illustris-1} cosmological simulation \cite{vogelsberger2014a,vogelsberger2014b,nelson2015} 
with a box of 106.5 comoving megaparsecs length and a baryonic mass resolution of $1.26\times{}10^6$~M$_\odot$. The 
population-synthesis simulations were run with {\sc mobse}, which includes state-of-the-art prescriptions for stellar 
winds and black hole formation \cite{giacobbo2018a,giacobbo2018b,giacobbo2019}. In our model, the BBH mass and 
merger rate strongly depend on the metallicity of progenitor stars; metal-poor ($Z\leq{}0.002$) massive stars are more 
efficient in producing merging BBHs and form more massive merging BBHs (up to approximatively $90$ M$_\odot$) than metal-rich 
stars \cite{giacobbo2018b}. The mass spectrum and merger rate of BBHs predicted from {\sc mobse} is fairly consistent 
with current constraints from GW detections \cite{abbottO2,abbottO2popandrate}. The {\sc mobse} simulation suit 
adopted here is named run "CC15$\alpha{}5$" and was already presented by \cite{giacobbo2018b}. For more details, we refer to \cite{mapelli2018} 
and \cite{giacobbo2018b}.

From the population-synthesis simulations, we generate catalogs of merging BBHs, which we plant in the cosmological 
simulations via a Monte Carlo algorithm (based on the star formation rate and metallicity evolution). From this procedure, we 
obtain a population of BBHs with astrophysically motivated mass and redshift distribution \cite{mapelli2019}.
In these simulations, we consider only BBHs formed from isolated binaries, and we neglect other possible formation 
channels (e.g., dynamical evolution in dense stellar systems \cite{PortegiesZwart:1999nm,Mapelli:2016vca,Rodriguez:2016kxx,Askar:2016jwt}.

Spins are assigned to BBHs {\it a posteriori}, by randomly drawing the dimensionless spin magnitude $\chi$ from a 
Maxwellian distribution with root mean square equal to 0.1. Spin orientations are assumed to be aligned with the binary 
orbital plane. 
The spin model is just a simple toy 
model because the astrophysical processes which affect the spin magnitude of black holes are still largely uncertain 
(Bouffanais \emph{et al.}, in preparation). Our simplified description of spins is consistent with current GW data \cite{abbottO2popandrate}.  

\begin{figure}[]
	\centering
	\includegraphics[width=0.5\linewidth]{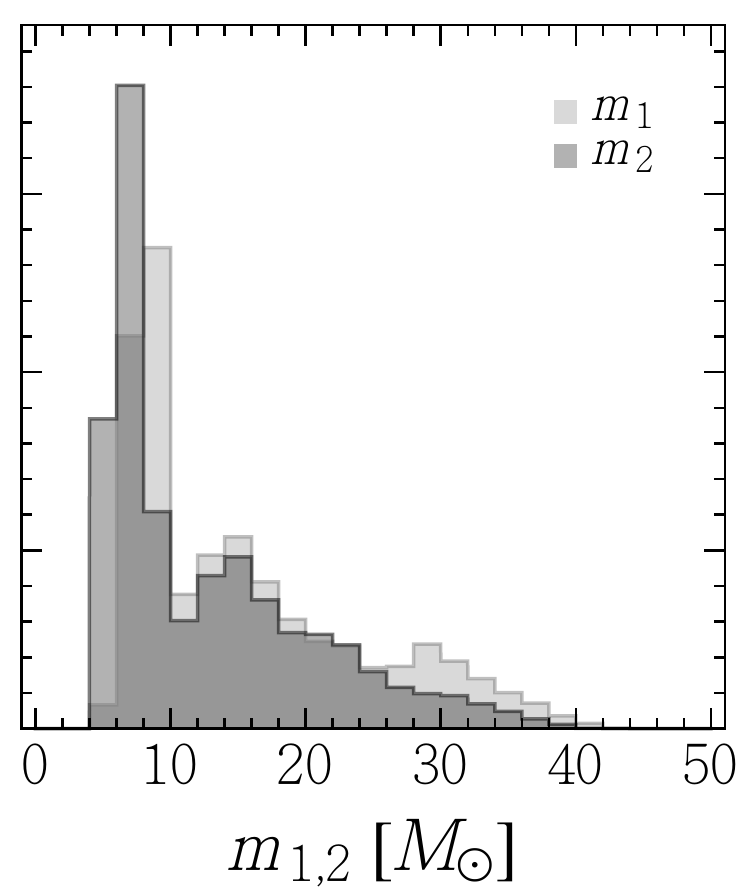}
	\caption[]{Distributions of component masses $m_{1}$ and $m_2$ for the black hole binaries considered in this paper.}
	\label{fig:masses}
\end{figure}

For this paper, we limit our sample to BBHs merging up to a luminosity distance of 1 Gpc (corresponding to redshift 
$z\sim{}0.2$), which is approximately the maximum distance at which LISA will be able to detect (rather massive) 
stellar BBHs \cite{Audley:2017drz,Isoyama:2018rjb}. The BBHs considered in this work span a mass ratio $1\lesssim m_1/m_2\lesssim 4$. 
The distribution of the component masses for the BBHs analyzed is shown in Fig.~\ref{fig:masses}.

\section{Waveform model}
\label{sec:waveform}
The main goal of our analysis is to compute bounds on alternative theories of gravity 
using multiband detections of the GW signals emitted by the BBH population described 
in Sec.~\ref{sec:stellarmodels}. As discussed in the Introduction, we work within the ppE 
framework, which introduces model-independent deviations from GR, both in the amplitude 
and in the phase of the gravitational waveform \cite{Yunes:2009ke}. In this paper, we focus 
on the inspiral part of the signals, taking into account non-GR corrections in the waveform 
phase only, as deviations in the GW amplitude can be considered subdominant~\cite{Cutler:1994ys,Barausse:2016eii}.
Extra information can be extracted from the other phases of a BBH coalescence, 
once the full inspiral-merger-ringdown emission in alternative theories is completely understood. 
New efforts have been currently devoted in this direction to study the binary evolution with 
GR modifications and fulfill the gap between analytical and numerical GW 
templates \cite{Hirschmann:2017psw,Okounkova:2017yby,Witek:2018dmd}.
 
Within the ppE approach, the template in the frequency domain is given by:
\begin{equation}
\label{eq:ppE_waveform}
h(f) = h_\mathrm{GR}(f) \, \mathrm{e}^{\mathrm{i}\beta u^b} \qquad  f \leq f_\mathrm{IM} \,,
\end{equation}
where $u = \pi \M f$ and ${\cal M}=(m_1m_2)^{3/5}/(m_1+m_2)^{1/5}$ is the chirp mass of the 
binary~\cite{Yunes:2009ke,Yunes:2013dva}. Note that $u \propto v^3$, where $v$ is the orbital 
velocity. For the GR waveform $h_\mathrm{GR}$, we use the PhenomB template for non-precessing, 
spinning BBHs~\cite{Ajith:2007kx,Ajith:2009bn}, assuming a Newtonian amplitude with average sky orientation.. 
The cutoff $f_\mathrm{IM}$ 
is the inspiral-merger transition frequency defined in~\cite{Ajith:2009bn}. Finally, $(b,\beta)$ are the 
ppE parameters, where $b$ controls the nature of the non-GR deviations and $\beta$ controls their magnitude. 
In the standard post-Newtonian terminology, a ppE coefficient identified by a given $b$ corresponds to a 
$(5+3b)/2$ PN order term. With a specific choice $b$, different modifications at 
various PN orders, including negative pre-Newtonian effects~\cite{Berti:2018cxi}, can be studied separately. 
In general, corrections with negative (positive) PN orders have a major impact on the low- (high-)frequency 
part of the GW signal. Such corrections are not completely unconstrained, since bounds in both the gravitational 
and the electromagnetic spectrum do exist~\cite{TheLIGOScientific:2016src,Yunes:2016jcc,Yunes:2010qb}.

\subsection{Non-GR modifications}
\label{sec:alternative}

We provide here a brief description of the three alternative theories of gravity considered in the 
paper. The non-GR modifications that such theories bring to the GW phase cover a wide range of
post-Newtonian orders. Therefore, they represent a good set of candidates to explore the feasibility 
of the multiband analysis, and to understand the complementarity of space-borne and ground-based 
interferometers 
to constrain a specific correction. We also focus on effects which generally belong to modifications 
of the GW generation mechanism~\cite{Yunes:2013dva}, which are active in the source's near zone, 
leading to changes in the binary equations of motion. The three alternatives considered fit the general 
class of scalar-tensor theories of gravity. These are among the most natural modifications of GR, 
in which the gravity sector is nonminimally coupled with an extra scalar field. 
The latter affects both the binary orbital evolution and the gravitational wave flux and induces  
the emission of a dipole radiation, which is forbidden in GR~\cite{1975ApJ...196L..59E}.
In these theories, the Klein-Gordon equation for the scalar field is in general sourced by two terms, 
$\square\phi\propto  S^\tn{matter}_1+ S^\tn{curv}_2$, which are proportional to the matter's stress-energy 
tensor and to the curvature corrections added to the Lagrangian, respectively. The presence of both or of 
only one of these two components depends on the specific theory considered and determines 
the features of the dipole emission.

The first and simplest GR extension we focus on is a {\it generalized} version of the 
Fierz-Jordan-Brans-Dicke scalar-tensor theory~\cite{Jenkins:2013fya}, in which the scalar field 
is coupled to the Ricci scalar. In this theory, $S_2^\tn{curv}=0$, and $\phi$ is sourced by the  
matter content only. This implies that for a globally vacuum spacetime, for which also $S_1^\tn{matter}=0$, 
the scalar-field profile is 
constant. Binary black holes therefore do not lead to a dipole emission\footnote{However, mixed 
binaries with at least one star do emit scalar dipole radiation.}. However, it is possible to have 
non-trivial configurations for $\phi$ for a specific choice of the boundary conditions~\cite{Berti:2013gfa}. We 
therefore follow the approach pursued in~\cite{Barausse:2016eii}, by considering a generic parametrization which 
captures the effect of Brans-Dicke-like (BD-like) theories through a single coefficient $B$, which quantifies the 
magnitude of the dipole flux in the GW luminosity, $\frac{dE}{dt} =\frac{dE_\tn{GR}}{dt} \sqbrackets{1+B\brackets{\frac{M}{l}}^{-1}}$, 
where $dE_\tn{GR}/dt$ is the quadrupolar GR component, $M=m_1+m_2$ is the total mass of 
the binary, and $l$ is its orbital separation.
BD-like theories are described within the ppE approach by the following choice of 
parameters,
\begin{align}
b_\text{BD-like} = -\frac{7}{3} \ ,\quad \beta_\text{BD-like}  = -\frac{3}{224}\nu^{2/5} B \, ,\label{eq:ppE_BD-like}
\end{align}
with $\nu = m_1 m_2/M^2$ denoting the symmetric mass ratio, and correspond to a $-1$PN correction, which is 
therefore dominant at low frequencies with respect to the standard emission in GR. 

We also consider two specific examples of quadratic gravity theories, i.e., Einstein-dilaton-Gauss-Bonnet\footnote{Note that 
the BD-like correction introduced before can also map EdGB modifications.} 
(EdGB)~\cite{Kanti:1995vq,Moura:2006pz} and dynamical Chern-Simons (dCS)~\cite{Jackiw:2003pm,Alexander:2009tp,Delsate:2014hba}, 
in which the Einstein-Hilbert action is modified by introducing more complex corrections, quadratic in the curvature, 
which couple with the scalar field $\phi$. In these theories both $S^\tn{matter}_1$ and $S^\tn{curv}_2$ are in 
general different from zero, and the matter-independent term leads to dipolar/quadrupolar 
emission in EdGB/dCS black hole binaries. 
Due to the nature of the non-GR terms, quadratic theories are particularly 
relevant in the strong-field regime of gravity. The ppE map to 
EdGB gravity leads to the following parameters~\cite{Yunes:2013dva}:
\begin{align}
b_\EdGB &= -\frac{7}{3}\nonumber\ ,\\ 
\beta_\EdGB&=- \frac{5\brackets{m^2_1 s^\GB_2 - m^2_2 s^\GB_1}^2}{7168 \, \nu^{18/5}M^4} \zeta_\EdGB \, .\label{eq:ppE_EdGB}
\end{align}
The phase amplitude $\beta_\EdGB$ depends on the coefficients $s^\GB_{i=1,2}$,
\begin{equation}
	s^\GB_i=2\frac{(\sqrt{1-\chi^2_i}-1+\chi_i^2)}{\chi_i^2} \, ,
\end{equation}
which show a strong correlation between the BH masses and the spin parameters 
$\chi_i\in[-1,1]$~\cite{Yunes:2016jcc}.
The value $b_\EdGB=-7/3$ identifies again a $-1$PN correction: BH binaries emit 
scalar dipole radiation induced by the black hole's individual monopole charges. 
For a given $\alpha_\EdGB$, which is the actual parameter entering the theory's Lagrangian and 
has the dimensions of a squared length, the dimensionless coupling $\zeta_\EdGB = (16\pi/M^4)\alpha^2_\EdGB$ 
is larger for low-mass sources. In this regard, the stellar-mass population studied in this work represents 
the ideal arena to test EdGB theory.

Finally, the last theory considered in this paper, dCS gravity, breaks two GR foundations. As for EdGB, 
dynamical Chern-Simons violates the strong equivalence principle. Moreover, the scalar field's equation 
is not invariant under parity transformation, due to the specific nature of the quadratic curvature correction, 
given by the Pontryagin invariant~\cite{Stein:2014wza}. However, unlike the theories described above, dCS introduces 
a 2PN deformation on the GR phase, specified by the following parameters,
\begin{align}
b_\dCS =&-\frac{1}{3}\, , \nonumber \\ 
\beta_\dCS = & \frac{481525}{3670016 \, \nu^{14/5}}\left[\brackets{1-\frac{67732}{19261}\nu}\chi_s^2  \right.\nonumber\\
&\left. +\brackets{1-\frac{9312}{19261}\nu}\chi_a^2 - 2\delta\chi_s \chi_a\right] \zeta_\dCS \, ,\label{eq:ppE_dCS}
\end{align}
where $\delta=(m_1-m_2)/M$, $\chi_s=(\chi_1+\chi_2)/2$, $\chi_a=(\chi_1-\chi_2)/2$, and 
$\zeta_\dCS = (16\pi/M^4)\alpha^2_\dCS$ \cite{Nair:2019iur}. In this theory the black holes feature a 
scalar dipole charge, which in turn induces a scalar quadrupolar emission during the orbital evolution.

\section{Statistical analysis }
\label{sec:analysis}

In this paper, we consider the sensitivity of four different detectors which form a multiband GW network: i) a LIGO/Virgo 
second generation interferometer at design sensitivity~\cite{LIGOWhite}; ii) the future third 
generation interferometer Einstein Telescope (ET) in the so-called ET-D configuration~\cite{Hild:2010id,ETWhite}; iii) 
the Japanese space-based interferometer B-DECIGO, also planned to test the 
capabilities of the larger detector DECIGO~\cite{Sato:2017dkf,Isoyama:2018rjb}; and iv) the space-borne detector 
LISA, with and Optical Measurement System\footnote{A more conservative value of the OMS, i.e. 15 pm/$\sqrt{\tn{Hz}}$, would reduce the signal-to-noise ratio 
of roughly $50\%$, increasing the errors of the same order of magnitude.} (OMS) of 10 pm/$\sqrt{\tn{Hz}}$ at high 
frequencies and assuming a mission lifetime of four years \cite{Audley:2017drz}. 

\begin{figure}[]
	\centering
	\includegraphics[width=1\linewidth]{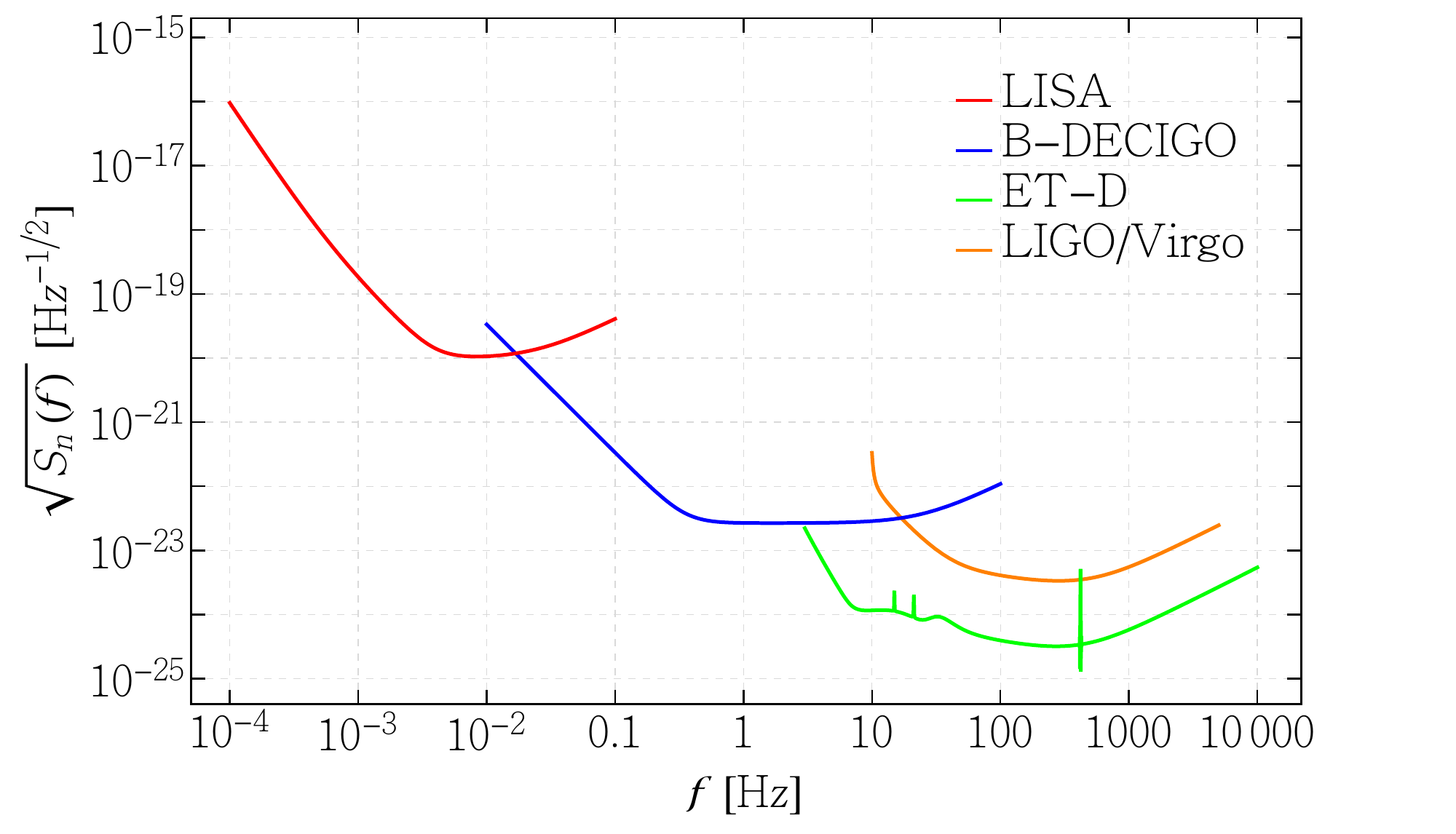}
	\caption{Power spectral density $S_n(f)$ as a function of the frequency for the four GW detectors 
	considered in the multiband network.
	\label{fig:sensitivity_curves}}
\end{figure}

The sensitivity 
curves of these four GW detectors are shown in Fig.~\ref{fig:sensitivity_curves}.

The gap of approximately $[0.1,3]\,\mathrm{Hz}$ between LISA and the ET can be filled by the decihertz band of 
B-DECIGO, which becomes, in this regard, crucial to construct a complete network of detectors with 
a seamless spectrum $f\in[10^{-4},10^{4}]\,\mathrm{Hz}$, as shown in Fig.~\ref{fig:sensitivity_curves}.

The estimate of the source parameters, and hence of the non-GR modifications, would greatly benefit from 
multiband observations~\cite{Barausse:2016eii}, especially for those theories which predict changes  
which are dominant in a specific part of the spectrum. For example, dipole radiation arising in EdGB or 
BD-like theories occurs at low frequencies as a pre-Newtonian correction, and it is expected to be 
constrained with more precision by space detectors operating at subhertz frequencies, like LISA and 
B-DECIGO. On the other hand, dCS gravity introduces a 2PN order effect, which becomes relevant 
at high frequencies, where ET plays a major role. 

\subsection{Signal-to-noise ratios and the Fisher matrix}
\label{sec:fisher}
In order to describe the mathematical tools used to compute the uncertainties on the ppE 
parameters, it is useful to introduce the inner product between two functions $A(t)$ and $B(t)$ 
in the waveform space, weighted on the power spectral density $S_n(f)$ of a given 
detector~\cite{Cutler:1994ys},
\begin{equation}
\label{eq:inner_product}
  \brackets{A|B}  = 4\mathrm{Re} \int_{f_\mathrm{in}}^{f_\mathrm{fin}}\frac{\tilde{A}(f)\tilde{B}^*(f)}{S_n(f)} \, df \,,
\end{equation}
with $\tilde{A}(f)$ and $\tilde{B}(f)$ being their Fourier transforms. The signal-to-noise ratio 
(SNR) $\rho$ of a GW signal $\tilde{h}(f)$ is then given:
\begin{equation}
\label{eq:def_SNR}
	\rho^2 = \brackets{h|h} = 4\int_{f_\mathrm{in}}^{f_\mathrm{fin}} \frac{|\tilde{h}(f)|^2}{S_n(f)} \ df \,.
\end{equation}
The bandwidth $[f_\mathrm{in},f_\mathrm{fin}]$ is fixed by the specific instrument and/or 
GW event considered. In particular, $f_\mathrm{fin}$ is chosen to be equal to $f_{\mathrm{IM}}$ 
when we consider LIGO/Virgo and ET and equal to 0.1 and 100 Hz for LISA and B-DECIGO, 
respectively.

On the other edge of the frequency spectrum, $f_\tn{min}$ is limited by seismic noise for 
ground-based detectors, $f^\tn{LIGO/Virgo}_\tn{min}=10$ Hz, $f^\tn{ET}_\tn{min}=3$ Hz, 
while we fix  $f_\tn{min}=0.01$ Hz for B-DECIGO~\cite{Isoyama:2018rjb}. Finally, 
following~\cite{Berti:2004bd}, the initial frequency for LISA is determined by:
\begin{equation}
	f_\mathrm{in} = 4.149 \times 10^{-5} \brackets{\frac{\M}{10^6 M_\odot}}^{-5/8} \,
	\brackets{\frac{\mathcal{T}_\tn{obs}}{1\mathrm{yr}}}^{-3/8}\mathrm{Hz}\ ,
\end{equation}
where $\mathcal{T}_\tn{obs}=4\,\mathrm{yr}$.

\begin{figure}[]
	\centering
	\includegraphics[width=0.6\linewidth]{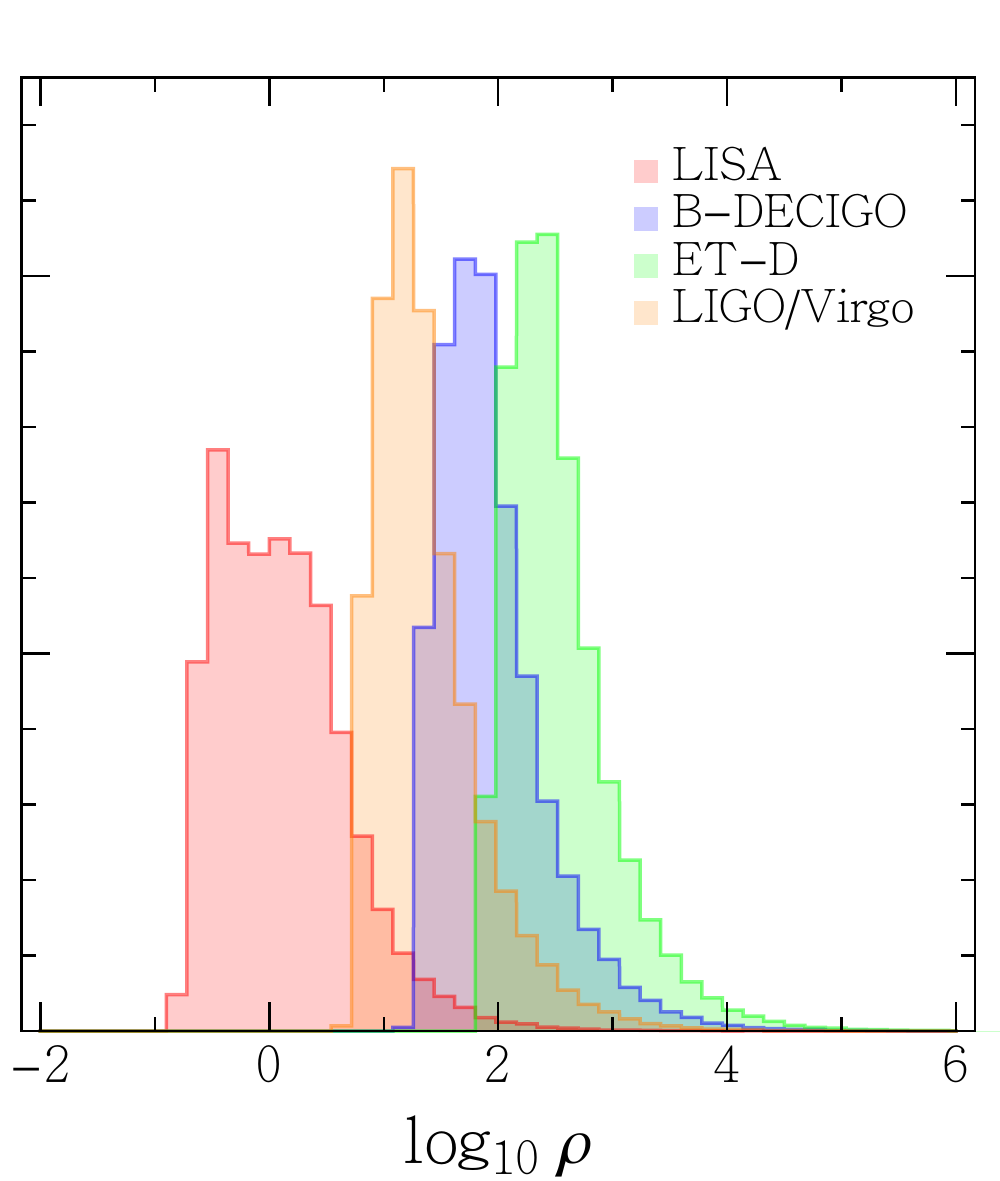}
	\caption{SNR of the BBH population described in Sec.~\ref{sec:stellarmodels} for the four detectors 
	of the multiband network. The SNR thresholds applied to the interferometers cut approximately $95\%$ 
	and $10\%$ of events in the LISA and in the LIGO/Virgo bands, respectively.}
	\label{fig:snrdetectors}
\end{figure}

In our analysis we focus on high-SNR signals, asking that the value of $\rho$ in each 
detector of the network is larger than a  fixed threshold, namely $\rho_\tn{th}=15$ for LISA 
\cite{Moore:2019pke} and $\rho_\tn{th}=8$ for the other interferometers~\cite{Abbott:2016nhf}. 
Figure~\ref{fig:snrdetectors} shows the SNR distributions for 
the population of binaries studied in this work. The requirement $\rho>\rho_\tn{th}$ 
cuts a large fraction of events observed by LISA. This is somehow 
expected, as the BBH analyzed feature a stellar-mass distribution, and therefore represent the 
prototype target for ground-based detectors, like ET. The latter shows the largest values of $\rho$, 
followed by B-DECIGO, for which the binaries accumulate SNR due to the long sweep in the 
low-frequency band.

Under the large-SNR hypothesis~\cite{Vallisneri:2007ev}, we can assume that the GW parameters 
determined by the data analysis are Gaussian distributed around the true values $\bar{\theta}$, 
i.e., $\theta=\bar{\theta}+\delta{\theta}$. In this case, assuming flat priors, the posterior probabilities 
for the source parameters can be written as~\cite{Cutler:1994ys},
\begin{equation}
\label{eq:def_fisher_pdf}
	\mathcal{P}(\theta) \propto  \mathrm{exp}\sqbrackets{ -\frac{1}{2}\,\Gamma_{ij}\,\delta\theta^i\delta\theta^j } \,,
\end{equation}
where $\Gamma_{ij}$ is the Fisher information matrix,
\begin{equation}
\label{eq:def_fisher}
	\Gamma_{ij}=\left(\frac{\partial h}{\partial\theta^i} \Bigg\vert \frac{\partial h}{\partial\theta^j}\right) \Bigg{|}_{\theta=\bar\theta} \,,
\end{equation}
with $(\,\cdot\,|\,\cdot\,)$ being the noise-weighted inner product~\eqref{eq:inner_product}.
In this framework, the covariance matrix is simply determined by the inverse of the Fisher matrix, 
$\Sigma_{ij} = (\Gamma^{-1})_{ij}$, and the parameter's standard deviation given by 
$\sigma_i  = \sqrt{ \Sigma_{ii}}$.
In our analysis, we consider $\theta = \curbrackets{\M,\nu,\chi,t_\mathrm{c},\phi_\mathrm{c},\theta_\ppE}$, 
where $\chi$ is the PhenomB effective spin parameter; $t_\mathrm{c}$ 
and $\phi_\mathrm{c}$ are, respectively, the time and phase at the coalescence; and $\theta_\ppE$ represents the 
ppE parameter that encodes the magnitude of the non-GR modification to the GW phase [cf. 
Eqs.~\eqref{eq:ppE_BD-like},~\eqref{eq:ppE_EdGB} and~\eqref{eq:ppE_dCS}].

We compute the Fisher matrix in each of the network's detectors for all the BBHs included in our catalogue, injecting 
the ``true'' values of the parameters $\bar\theta$ in Eq.~\eqref{eq:def_fisher} under the null hypothesis that 
GR is the correct theory of gravity, i.e., such that $\bar\theta_\ppE = 0$. Assuming that the GW observations 
are independent, it is straightforward to combine different datasets for LIGO/Virgo, ET, B-DECIGO and LISA, 
in order to obtain multiband constraints. In this case,
\begin{equation}
	\sigma_i^{2\,\tn{tot}} = \left(\sum_{k=1}^{N_d}\Gamma_{(k)}\right)^{-1}_{ii} \ ,\label{multfisher}
\end{equation}
where the index $k$ runs over the $N_d$ detectors considered and $\Gamma_{(k)}$ are the associated 
Fisher matrices. Note that for network configurations, with LISA and one ground-based detector, 
eq.~\eqref{multfisher} is strictly valid only for signals which are phase connected \cite{Barausse:2016eii}. 
The results presented in the next section for the multiband analysis correspond to the BBHs 
which satisfy this requirement.

\section{Results}
\label{sec:results}

Assuming that GR is correct, the standard deviation of each ppE parameter resulting from the Fisher 
matrix analysis imposes an upper bound on the corresponding modified theory of gravity at $1\sigma$ 
confidence level.

\begin{figure*}[]
	\centering
	\includegraphics[width=0.25\linewidth]{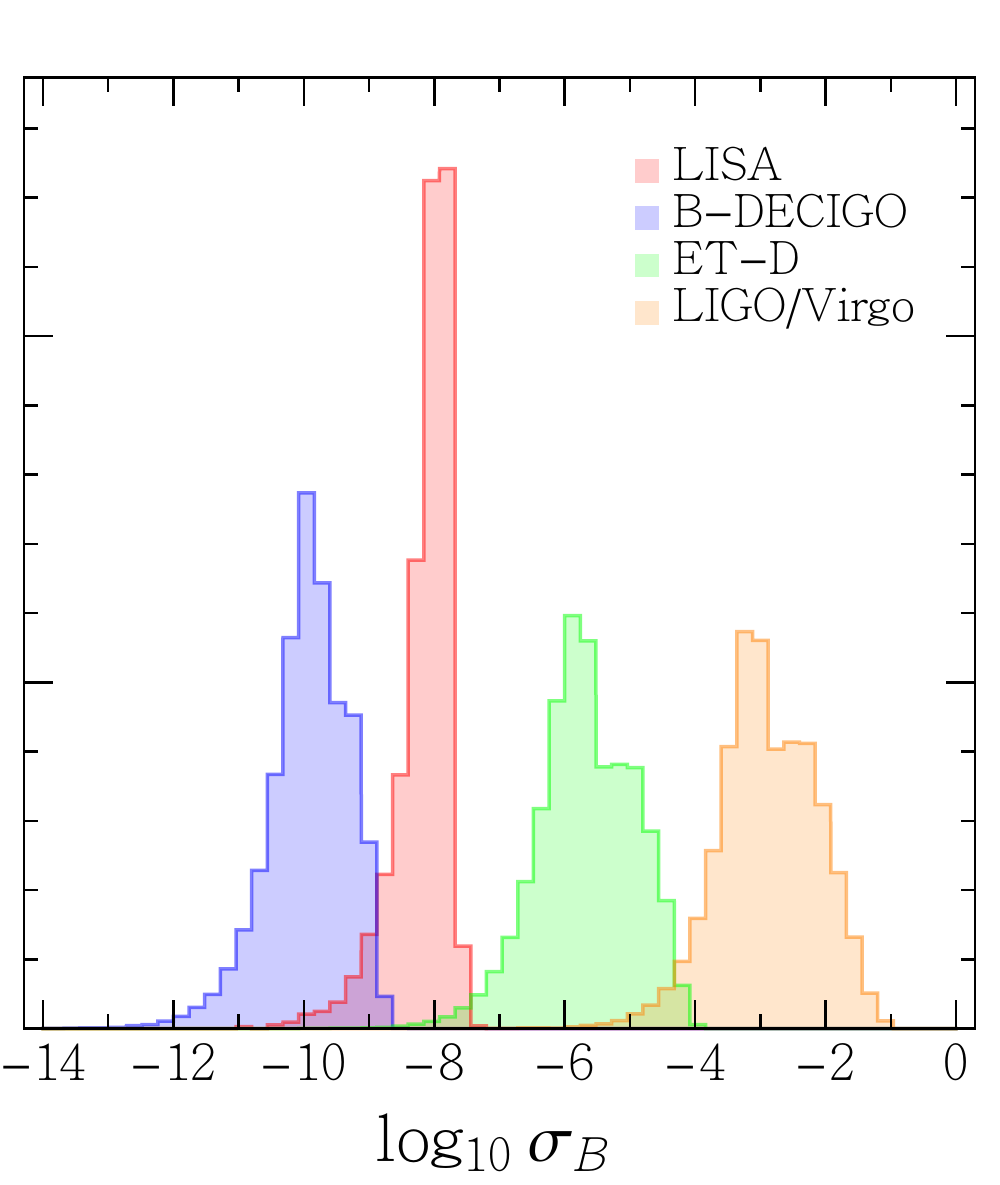}
	\includegraphics[width=0.25\linewidth]{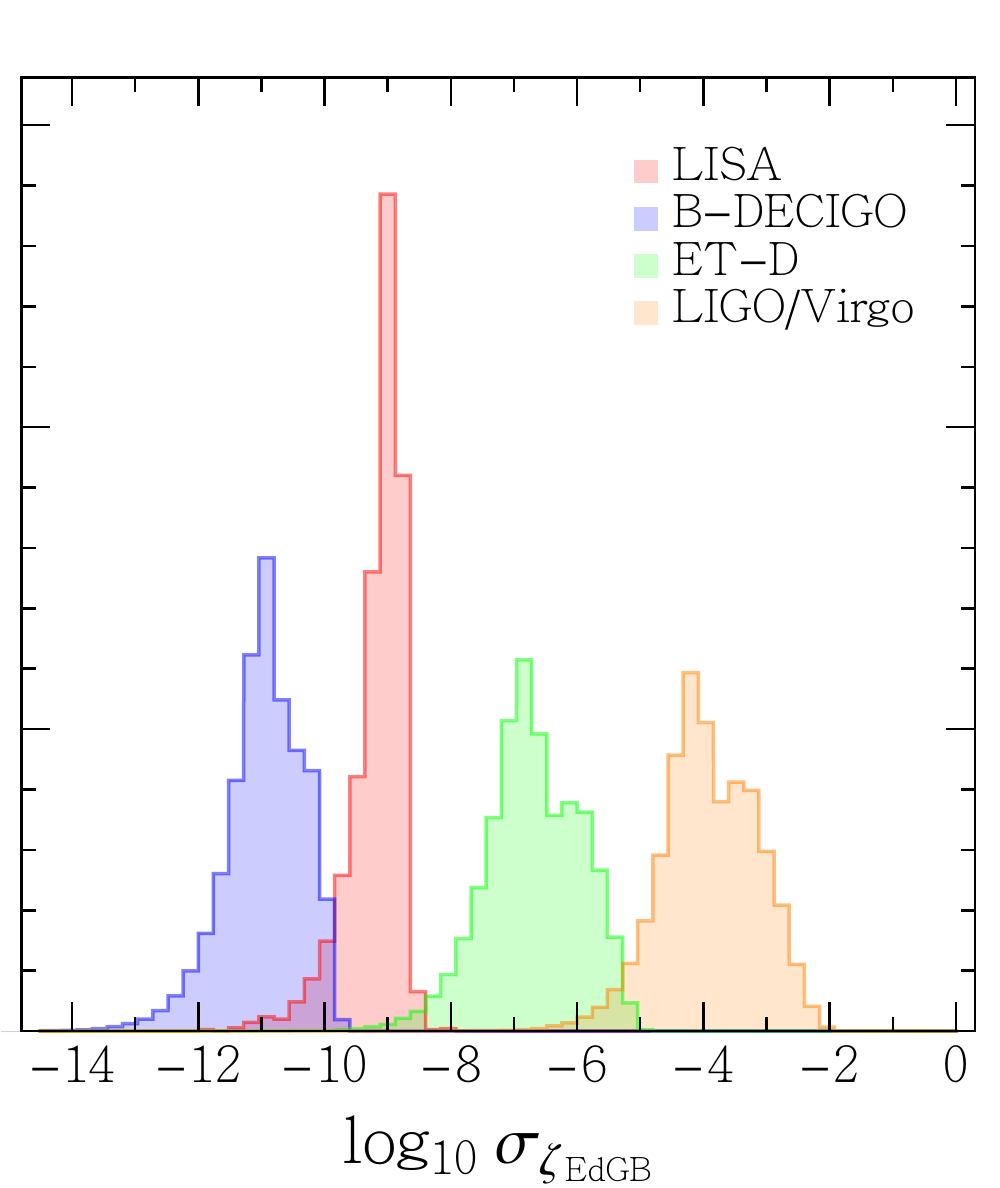}
    	\includegraphics[width=0.25\linewidth]{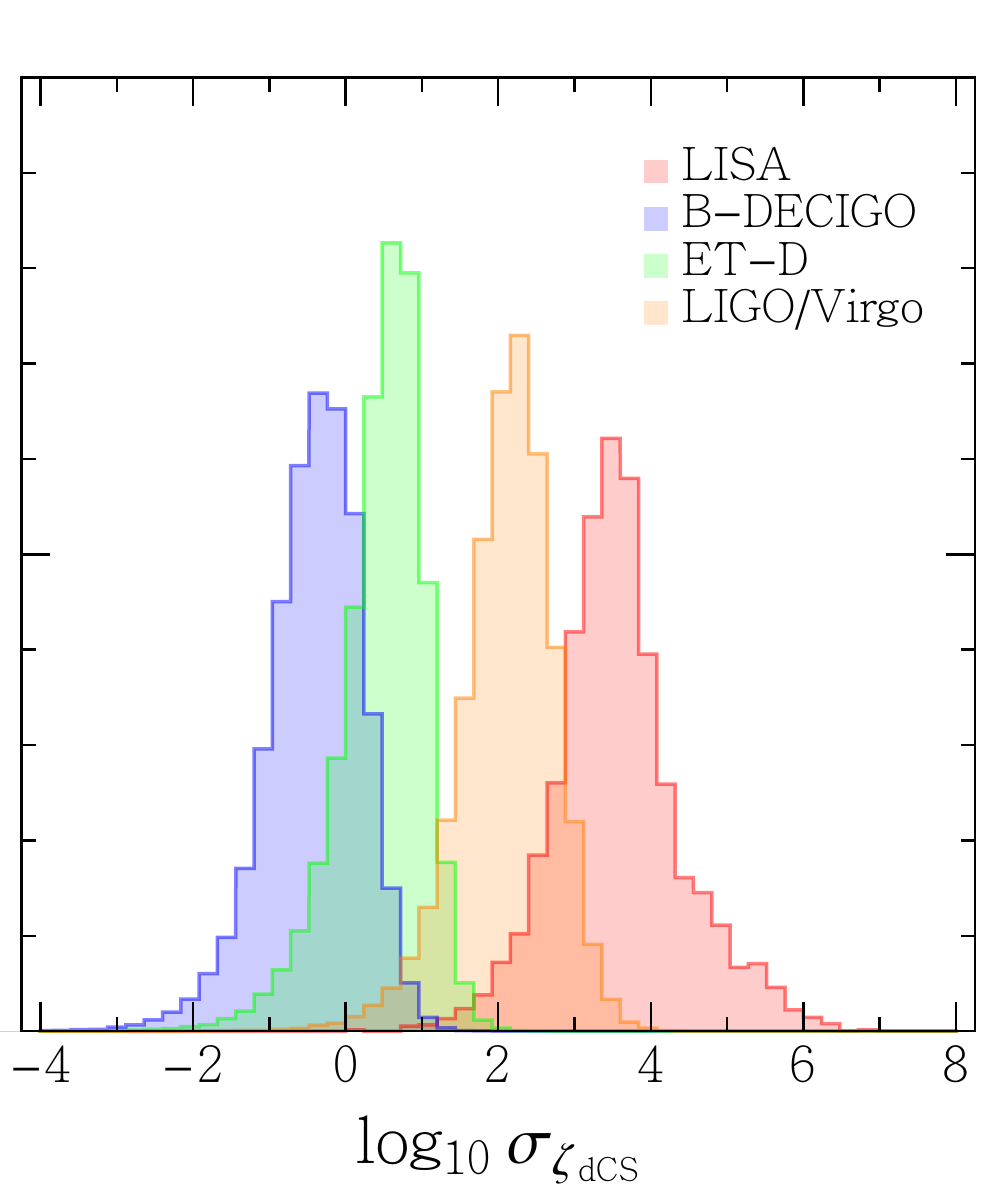}
	\caption[]{Distributions of the $1\sigma$ upper bounds on the parameter $B$ which controls the dipole radiation flux 
	in BD-like theories (left panel), on the coupling constants of EdGB (center panel) and of dCS (right panel). Different 
	colors refer to measurements obtained by distinct detectors. We only consider BBH events for which the SNR 
	satisfies the condition $\rho\ge\rho_\tn{th}$ (see Sec.~\ref{sec:fisher}).}
	\label{fig:sigma}
\end{figure*}

Figure~\ref{fig:sigma} shows the distributions of the $1\sigma$ errors for the parameters which characterize 
the three alternative theories described in Sec.~\ref{sec:alternative} and the BBH population of 
Sec.~\ref{sec:stellarmodels}. For each detector, we only consider sources yielding a SNR above the 
threshold $\rho_\tn{th}$. This requirement excludes approximately $90\%$ of the binaries in the LISA band, 
while it has small impact for the other detectors. 
As expected, for both BD-like and EdGB, which predict a $-1$PN correction to the GW phase, low-frequency detectors 
provide the strongest constraints. 
For the two space detectors, we obtain: $B \lesssim 10^{-10}$ and 
$\zeta_\mathrm{EdGB}\lesssim 10^{-11}$ for B-DECIGO, whereas $B \lesssim 10^{-8}$ and 
$\zeta_\mathrm{EdGB}\lesssim10^{-9}$ for LISA ($1\sigma$ median values; see Table~\ref{tab:medianvalues}).
The most stringent constraints on $B$ and on the EdGB coupling constant come from the Japanese detector, which 
compared to LISA features higher values of the SNR. In the EdGB case, it is also crucial 
to have very accurate measurements of the BH spin parameters \cite{Yunes:2016jcc}, 
which are correlated with $\zeta_\mathrm{EdGB}$ (and the masses) in the GW phase 
through the ppE coefficient~\eqref{eq:ppE_EdGB}. 
In this regard, we note that B-DECIGO is able to perform measurements 
of $\chi$ with a relative accuracy well below 1\% for all the considered BH binaries, compared against 
the errors obtained by LISA, which cluster around $\sigma_\chi/\chi\sim 5\%$.

\begin{table}[]
\centering
	\renewcommand*{\arraystretch}{1.3}
\setlength{\tabcolsep}{7pt}
    \begin{tabular}{lcccc}
	\hline
    \hline
   & LISA & B-DEC & ET & LV \\
 	\hline
$\sigma_B$ 		                      & $9.0\cdot10^{-9}$  & $1.3\cdot 10^{-10}$ & $2.0\cdot10^{-6}$  & $1.3\cdot10^{-3}$\\
$\sigma_{\zeta_\mathrm{EdGB}}$        & $8.7\cdot10^{-10}$ & $1.2\cdot10^{-11}$  & $1.9\cdot10^{-7}$  & $1.2\cdot10^{-4}$\\
$\sigma_{\zeta_\mathrm{dCS}}$         & $502$ 			   & $4.5\cdot 10^{-1}$  & $3.6$  			  & $146$  			 \\
   \hline
$\sigma_{\alpha_\mathrm{EdGB}^{1/2}}$ & $1.4\cdot10^{-1}$  & $2.5\cdot10^{-2}$ 	 & $2.9\cdot10^{-1}$  & $1.7$			 \\ 
$\sigma_{\alpha_\mathrm{dCS}^{1/2}}$  & $114$ 			   & $12$ 				 & $19$ 			  & $53$ 			 \\ 
   \hline
      \hline
  \end{tabular} 
  \caption{(Top rows) Median values of the distributions in Figs.~\ref{fig:sigma} and~\ref{fig:sigma_multiband_PDF} 
  for the 68\% confidence level bounds on the alternative theories described in 
  Sec.~\ref{sec:alternative} measured with LISA, B-DECIGO, ET and advanced detectors (LV).
  (Bottom rows) Median values of the distributions  for the 95\% $(2\sigma)$ confidence level bounds 
  on the coupling constants (in kilometers) of EdGB and dCS theories.}\label{tab:medianvalues}
\end{table}

These results change when we analyze the data in terms of ground-based detectors. The left and center panels 
of Fig.~\ref{fig:sigma} show indeed that both third and second  generation interferometers would constrain BD-like 
and EdGB parameters several orders of magnitude less than space-borne detectors. For ET, in the best case 
scenario, the new bounds on $\zeta_\mathrm{EdGB}$ can be as large as 6 (7) orders of magnitude with respect 
to B-DECIGO (LISA). This is mainly due to the sensitivity in the frequency band of space-borne instruments with respect to the 
dipole correction and to the ability to constrain the BH spins with better accuracy. For BD-like 
corrections our results are consistent with the uncertainties on $B$ found in \cite{Barausse:2016eii,Chamberlain:2017fjl} 
for stellar-mass binaries observed by both ground-based and space interferometers.

The right plot of Fig.~\ref{fig:sigma} shows the distributions of projected bounds for dCS theory. As explained in 
Sec.~\ref{sec:alternative}, in this case, the non-GR modifications lead a  2PN correction in the GW signal, which is  
therefore expected to contribute more at higher frequencies. The constraints coming from LISA are much looser and 
compatible with the results of advanced detectors, while ET is able to put more stringent bounds and 
to measure $\zeta_\mathrm{dCS} \lesssim 1$ ($1\sigma$ median value). Note that the values obtained for LISA and 
LIGO/Virgo violate the small-coupling approximation, namely $\zeta_{\mathrm{EdGB,dCS}}\ll 1$, assumed to derive eq.~\eqref{eq:ppE_dCS}.
Even for dCS, the detector which provides the tightest bounds is B-DECIGO, for which we obtain a median 
$\zeta_\mathrm{dCS} \lesssim 10^{-1}$. This result is again due to the exquisite sensitivity of 
B-DECIGO to the inspiral phase of stellar-mass objects, which is able to constrain with good accuracy all 
the waveform parameters. Comparing the three panels of Fig.~\ref{fig:sigma}, we finally note that the constraints 
on GW deviations arising from BD-like and EdGB are in general several orders of magnitude stronger than those 
on the dCS theory. This difference is mainly related to the different PN order which characterizes the non-GR 
modifications in the gravitational waveform.

For both quadratic gravity theories, knowing the mass of the BBHs and their uncertainties we can propagate the 
errors on $(\zeta_\mathrm{EdGB},\zeta_\mathrm{dCS})$ into bounds on $(\alpha_\mathrm{EdGB},\alpha_\mathrm{dCS})$, 
which are the fundamental parameters entering the EdGB and dCS actions. In last two rows of Table~\ref{tab:medianvalues}, 
we show the median values of the $2\sigma$ upper limits obtained for $(\alpha_\mathrm{EdGB}^{1/2},\alpha_\mathrm{dCS}^{1/2})$ 
by propagating the measurement uncertainties. The conversion reduces the difference in magnitude among the constraints imposed 
by each interferometer. The best results, yielded by B-DECIGO, are $\alpha_\mathrm{EdGB}^{1/2} \lesssim 0.03$ km 
and $\alpha_\mathrm{dCS}^{1/2} \lesssim 12$ km.\\

\begin{figure}[]
	\centering
	\includegraphics[width=0.65\linewidth]{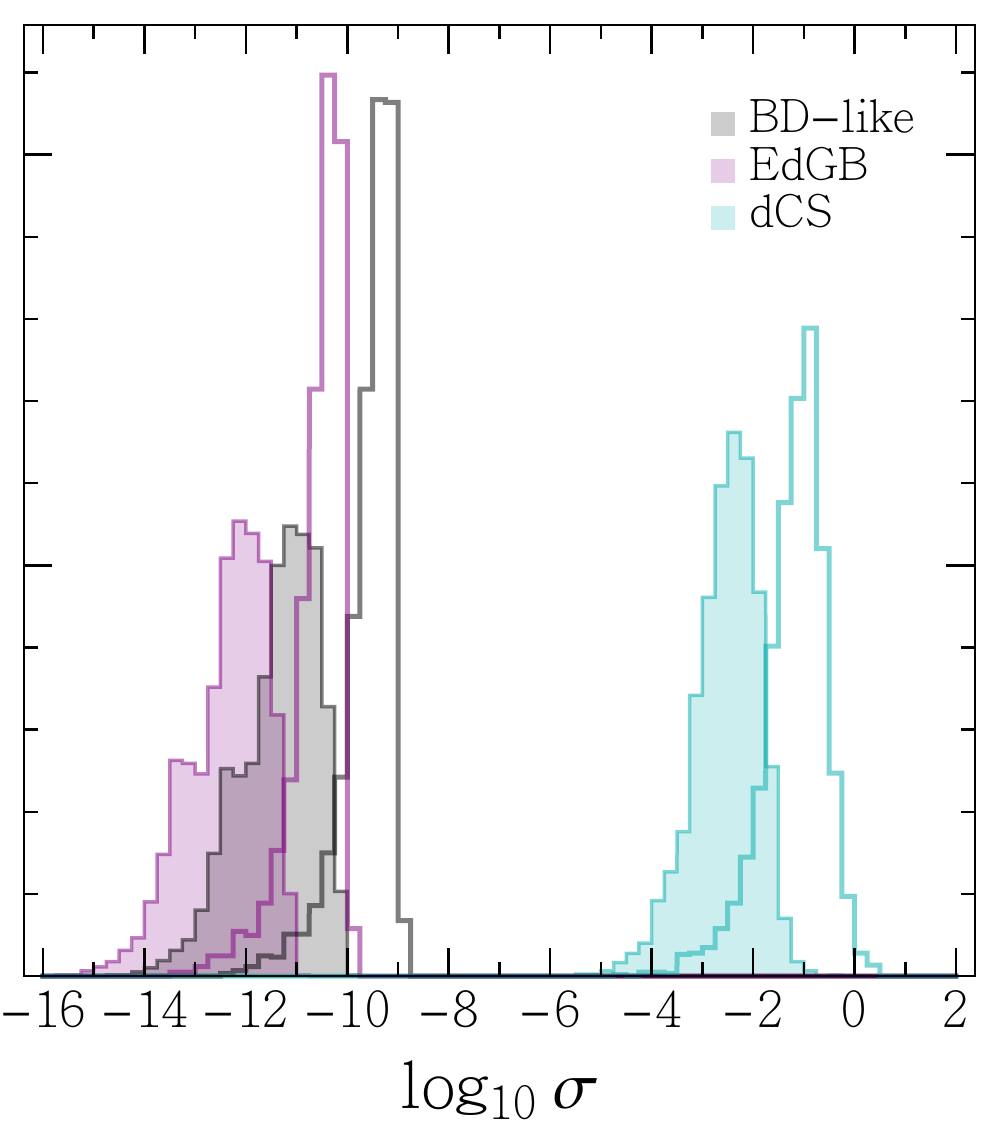}
	\caption[]{Probability distributions of the $1\sigma$ upper bounds on the parameters of the three alternative 
	theories of gravity considered, using multiple detectors. Filled and empty histograms refer to errors computed 
	with and without B-DECIGO, respectively.}
	\label{fig:sigma_multiband_PDF}
\end{figure}

Having determined the statistical errors obtained in the single-detector scenario, we can now analyze how 
these results change when multiple detectors are taken into account. 
Note that in this case we are assuming that the BBHs are observed simultaneously by all the detectors. 
This naturally restricts the total number of events to the fraction of binaries detected by LISA which, given the 
threshold requirement $\rho>15$, is approximately $5\%$ of the full population. Therefore, the 
ensemble of BBHs observed by B-DECIGO, ET, and LV will also be smaller compared to that one analyzed for 
the stand-alone configuration.
Figure~\ref{fig:sigma_multiband_PDF} 
shows the joint projected constraints derived using LISA+ET+LIGO/Virgo  (empty histograms) and the full network 
of four detectors (filled histograms), i.e., including B-DECIGO. The median values of the $1\sigma$ upper bounds are 
listed in Table~\ref{tab:medianvalues2}. For sake of clarity, we also show the constraints for single interferometer, 
assuming the restricted BBHs ensemble used for the multiband analysis.
The latter improves the results achieved by the single detectors, providing stronger bounds. When considering 
the full network, the errors derived for the dipole correction are very close to those obtained by B-DECIGO alone, 
which dominates the other interferometers. However, if we neglect the Japanese space detector, the joint 
LISA-ET-LV analysis  improves the bounds on $\sigma_B$ and $\sigma_{\zeta_\tn{EdGB}}$ by  more than 
a factor 20 with respect to the smallest constraints given by LISA. For dCS, the Einstein Telescope 
plays a more fundamental role for both the network configurations with and without 
B-DECIGO. Finally, constraints derived for the coupling constants $(\alpha_\mathrm{EdGB},\alpha_\mathrm{dCS})$ 
are also shown in the bottom rows of Table~\ref{tab:medianvalues2}.
\begin{table*}[]
\centering
	\renewcommand*{\arraystretch}{1.3}
\setlength{\tabcolsep}{7pt}
    \begin{tabular}{lcccccc}
	\hline
    \hline
    & LISA & B-DECIGO & ET-D & LVC & LISA+ET+LV & Full network\\
 	\hline
$\sigma_B$    						  
	& $9.0\cdot10^{-9}$ & $1.9\cdot10^{-11}$ & $5.1\cdot10^{-7}$ & $2.7\cdot10^{-4}$ & $3.6\cdot10^{-10}$ & $5.8\cdot10^{-12}$\\
$\sigma_{\zeta_\mathrm{EdGB}}$ 	 	  
	& $8.7\cdot10^{-10}$& $1.9\cdot10^{-12}$ & $4.9\cdot10^{-8}$ & $2.6\cdot10^{-5}$ & $3.5\cdot10^{-11}$ & $5.8\cdot10^{-13}$\\
$\sigma_{\zeta_\mathrm{dCS}}$ 		  
	& $502$ 			& $1.7\cdot10^{-2}$  & $2.0\cdot10^{-1}$ & $13$ 			 & $8.0\cdot10^{-2}$  & $3.5\cdot10^{-3}$ \\
	\hline
$\sigma_{\alpha_\mathrm{EdGB}^{1/2}} \mathrm{ [km]}$
	& $1.4\cdot10^{-1}$ & $3.4\cdot10^{-2}$   & $4.3\cdot10^{-1}$ & $2.1$ 			 & $8.4\cdot10^{-2}$  & $2.5\cdot10^{-2}$ \\
$\sigma_{\alpha_\mathrm{dCS}^{1/2}} \mathrm{ [km]}$ 
	& $114$ 			& $10$			      & $19$ 			  & $54$ 	   		 & $18$     		  & $6.8$ \\
   \hline
   \hline
  \end{tabular} 
  \caption{Median values of the distributions for the 68\% confidence level bounds on the 
  GR modifications for the multiband analysis. Note that the BBH events analyzed in this 
  case are simultaneously observed by all the interferometers and therefore correspond to 
  a smaller sample than that considered in Table~\ref{tab:medianvalues}. 
  The last two rows show the median for the 95\% confidence level bounds on the EdGB and dCS 
  couplings.}\label{tab:medianvalues2}
\end{table*}
The multiband analysis leads to narrower upper limits on the parameters of both quadratic gravity 
theories, which are factors approximately $10^{-2}$ and $10^{-7}$ smaller than current bounds on EdGB and dCS, 
respectively \cite{Yunes:2016jcc,Nair:2019iur}. These results are also consistent with previous analyses 
carried out on the evolution of different types of compact binaries in quadratic gravity theories \cite{Yagi:2012vf,Yagi:2012gp,Yamada:2019zrb}.

\section{Conclusions}

Current and future observations of gravitational wave signals from BBHs carry the full 
potential of testing General Relativity and tracing modifications of gravity in the relativistic strong-field 
regime. The constraints on possible non-GR deviations benefit from the synergy of multiple detectors, 
which greatly improve the accuracy on the measurements of the source's parameters. While LIGO 
and Virgo are already working as a network of ground-based interferometers, a complementary window 
will be soon provided from the outer space by LISA. Moreover, third generation detectors 
like ET, or future satellites such as B-DECIGO, will further enlarge this web of instruments, increasing the 
accuracy of GW observations.
 
In this paper, we analyze how the different setup of a network composed of LISA, LIGO/Virgo, 
ET, and B-DECIGO will constrain the fundamental parameters of three scalar-tensor theories, 
which induce modifications in the waveform at various post-Newtonian 
orders. Specifically, we consider a generalization of the Brans-Dicke theory and two types of quadratic 
gravity. We perform a Fisher matrix analysis on the signals produced by a population of stellar-mass, 
spinning BH binaries. Assuming that GR is the correct theory of gravity, we derive the probability 
distribution of the upper bounds on the parameters of each alternative theory. 
Our analysis assesses the effectiveness of multiband gravitational wave observations to put stringent 
bounds on deviations from Einstein's predictions. Among all detectors, we find that B-DECIGO imposes the 
strongest constraints on all the theories considered. As expected, LISA and ground-based detectors are 
able to detect with good accuracy low- and high-frequency modifications of the binary inspiral, respectively. 

We then derive joint constraints for the full network of interferometers by combining the results of each instrument. 
When B-DECIGO is not taken into account, the multiband analysis improves the upper bounds on the GR 
deviations up to an order of magnitude with respect to the single detector configuration. However, the uncertainties 
tend to be dominated by the Japanese detector, when the latter is included in the analysis. 
For a science case, we therefore point out that B-DECIGO plays a fundamental role in testing GR modifications, 
leading to the tightest bounds on all the alternative theories investigated and therefore supporting the quest for an 
intermediate-frequency detector between LISA and the ground-based interferometers.

Overall, our results suggest that existing bounds on the three alternative theories of gravity considered in this 
paper can be improved by several orders of magnitude~\cite{Barausse:2016eii,Yunes:2016jcc,Yagi:2012gp,AliHaimoud:2011fw}, 
and that a web of GW detectors able to cover a frequency band within $[10^{-4},10^3]$ Hz, 
represents a crucial ingredient to achieve precise tests of GR in the strong gravity regime. 

\textit{Note Added -} 
Recently, a preprint with similar conclusions appeared as 
an e-print \cite{2019arXiv190513155C}.

\begin{acknowledgments}
We acknowledge Valeria Ferrari for all her valuable discussions and comments during the developing 
of this project. We also thank Francesco Pannarale, Enrico Barausse, Leonardo Gualtieri, 
Xisco Jimenez-Forteza and the LIGO/Virgo referee 
Damir Buskulic for having carefully read the manuscript. 
T.A. acknowledges financial support provided under the European
Union's H2020 ERC, Starting Grant agreement no. DarkGRA--757480. M.M. acknowledges financial support by 
the European Research Council for the ERC Consolidator grant DEMOBLACK, under contract no. 770017. We 
acknowledge support from the Amaldi Research Center funded by the MIUR program ``Dipartimento di 
Eccellenza'' (CUP: B81I18001170001). The authors would also like to acknowledge networking support by the COST Action CA16104
\end{acknowledgments}
\vspace{7pt}

\bibliographystyle{iopart-num}
\bibliography{biblio}

\end{document}